\documentclass[%
 aip,
 amsmath,amssymb,
preprint,
]{revtex4-1}

\usepackage{graphicx}
\usepackage{dcolumn}
\usepackage{bm}
\usepackage{amsthm}

\begin{document}
\theoremstyle{theorem}
\newtheorem{theorem}{Theorem}

\theoremstyle{definition}
\newtheorem*{definition}{Definition}
\newtheorem*{remark}{Remark}
\theoremstyle{proposition}
\newtheorem{proposition}{Proposition}
\newtheorem*{conjecture}{Conjecture}

\title[Peculiar polynomials]{
The peculiar (monic) polynomials, the zeros of which equal 
their coefficients}
\author{F. Calogero}
\affiliation{
Dipartimento di Fisica, Universit\`{a} di Roma ``La Sapienza'' and Istituto di Fisica
Nucleare, Sezione di Roma, Rome, Italy
}

\author{F. Leyvraz}
 \altaffiliation[Also at ]{Centro Internacional de Ciencias.}

\affiliation{ 
Instituto de Ciencias F\'isicas---UNAM, Av.~Universidad s/n, Cuernavaca, 62210, Morelos, M\'exico
}

\date{\today}

\begin{abstract}
We evaluate the number of monic polynomials (of arbitrary degree $N$) the zeros
of which equal their coefficients when these are allowed to take 
arbitrary complex values. In the following, we call polynomials with this
property {\em peculiar\/} polynomials. We further show that the problem of determining 
the peculiar polynomials of degree $N$ simplifies when any of the coefficients is either 0 or 1. 
We proceed to estimate the numbers of peculiar polynomials of degree $N$ having one 
coefficient zero, or one coefficient equal to one, or neither.  
\end{abstract}

\maketitle

\section{Introduction}

\label{sec:intro}

Over half a century ago S.M.~Ulam \cite{Ulam1960} discussed the
transformation from $N$ numbers $x_{n}$ to their $N$ symmetrical sums $%
\sigma _{m}\left( \tilde{x}\right) $ multiplied by $\left( -1\right) ^{m}$, 
\begin{subequations}
\label{UlamTransf}
\begin{eqnarray}
y_m&=&\left( -1\right) ^{m}~\sigma _{m}\left( \tilde{x}\right)
~,~~~m=1,...,N~,  \label{ym} \\
\sigma _{m}\left( \tilde{x}\right) &=&\sum_{1\leq n_{1}<n_{2}<...<n_{m}\leq
N}x_{n_{1}}x_{n_{2}}\cdot \cdot \cdot x_{n_{m}}~,  \label{sigmam}
\end{eqnarray}%
implying, of course, 
\begin{eqnarray}
\sigma _{1}(\tilde{x}) &=&x_{1}+\ldots +x_{N},  \label{sigma1} \\
\sigma _{N}(\tilde{x}) &=&x_{1}\cdot \ldots \cdot x_{N}.  \label{sigmaN}
\end{eqnarray}%
Here $\tilde{x}$ is the unordered set of the $N$ numbers $x_{n}$ (see below).

It is well-known that the inversion of this transformation, 
(\ref{UlamTransf}), corresponds to the identification of the $N$ zeros of a polynomial of
degree $N$ with coefficients $y_{m}$. While calling attention to this
transformation \textquotedblleft operating on the $N$-dimensional real space
or on the $N$-dimensional complex space\textquotedblright\ \cite{Ulam1960}%
---Ulam pointed out the interest of investigating its \textit{fixed points},
namely the identification of \textit{all} the monic polynomials of degree $N$
\textit{with zeros }$x_{n}$\textit{\ equal to their coefficients }$y_{m}$.
Soon afterwards this problem was completely solved in the \textit{real}
domain---when the $N$ polynomial coefficients $y_{m}$ and the $N$ polynomial
zeros $x_{n}$ are \textit{all real} numbers---by Paul R. Stein \cite%
{Stein1966}. He proved that in this \textit{real} context---and restricting
attention only to polynomials with \textit{all} zeros $x_{n}$, hence \textit{%
all} coefficients $y_{m}$, \textit{nonvanishing} (a well justified
restriction, see below)---there are \textit{no} polynomials of this kind
with degree $N>4$. He moreover identified \textit{all} the polynomials of
this kind with degree $N\leq 4$: for $N=2$ the single polynomial $\left(
z-1\right) \left( z+2\right) =z^{2}+z-2$ (already mentioned by Ulam \cite%
{Ulam1960}); for $N=3$ the two polynomials $\left( z-1\right) \left(
z+1\right) ^{2}=z^{3}+z^{2}-z-1$ and $z^{3}+y_{1}z^{2}+y_{2}z$ $+y_{3}$ with 
$y_{1}=w$, $y_{2}=1/w$, $y_{3}=1/\left( w+1\right) $ where $w\approx
.5651977$ is the single \textit{real} root of the cubic equation $%
2w^{3}+2w^{2}-1=0$; for $N=4$ the single polynomial $%
z^{4}+y_{1}z^{3}+y_{2}z^{2}$ $+y_{3}z+y_{4}$ with $y_{1}=1$, $y_{2}=w\approx
-1.7548777$, $y_{3}=1/w\approx -0.5698403$, $y_{4}=y_{1}y_{2}y_{3}=1,$ where 
$w$ is the \textit{single real} root of the third degree
equation $w^{3}+2w^{2}+w+1=0$.

In this paper we treat the same problem but in the more general context of 
\textit{complex} numbers. A motivation to do so is because this is generally
a more natural context to discuss properties of polynomials since the
complex numbers are algebraically closed. A second motivation is connected
with the recently introduced notion of \textit{generations} of (monic)
polynomials, characterized by the property that the coefficients of the
polynomials of degree $N$ of a generation coincide with (one of the $N!$
permutations of) the zeros of the polynomials of the previous generation,
see \cite{BC2016}. Earlier work on this subject includes, to the best
of our knowledge, a paper by Di Scala and Maci\'a \cite{DiScala} proving that 
for any given degree, only finitely 
many such polynomials exist, as well as a paper by Bihun and Fulghesu
\cite{Bihun} in which they independently
derive some of the results we present here and further show that,
among these polynomials,
only the polynomial $z^N$ is an eigenfunction of a hypergeometric operator.  

In the following Section \ref{sec:def} we review our notation and
terminology and in Section \ref{sec:results} we report our main findings,
which are then proven in Section \ref{sec:proofs}. We present conclusions in 
Section \ref{sec:conclusions}.

\section{Notation and definitions}
\label{sec:def}

\subsection{Notation}

Hereafter we suppose to work---unless otherwise indicated ---with \textit{%
complex} numbers and with \textit{monic} polynomials of (\textit{positive
integer}) degree $N\geq 2$ in the \textit{complex} variable $z$,

\end{subequations}
\begin{equation}
p_{N}\left( z;\vec{y},\tilde{x}\right)
=z^{N}+\sum_{m=1}^{N}y_{m}~z^{N-m}=\prod\limits_{n=1}^{N}\left(
z-x_{n}\right) ~,  \label{pN}
\end{equation}%
which are characterized by their $N$ coefficients $y_{m}$ (the $N$
components of the $N$-vector $\vec{y}$) and by their $N$ zeros $x_{n}$ (the $%
N$ elements of the unordered set $\tilde{x}$). Of course above and hereafter
indices such as $n,$ $m$ are \textit{positive integers} in the range from $1$
to $N$ (unless otherwise indicated). The notation $p_{N}\left( z;\vec{y},%
\tilde{x}\right) $, see (\ref{pN}), is somewhat redundant, since this monic
polynomial is equally well identified by assigning {\em either\/}
its $N$ coefficients $y_{m}$ \textit{or} its $N$ zeros $x_{n}.$ Indeed its $N
$ coefficients $y_{m}$ are \textit{explicitly} expressed in terms of its $N$
zeros by the well-known formulas (\ref{UlamTransf}). Conversely, the \textit{%
unordered} set $\tilde{x}$ of the $N$ zeros $x_{n}$ of the polynomial (\ref%
{pN}) is uniquely identified when the $N$ coefficients $y_{m}$ of this
polynomial are assigned, although of course \textit{explicit} formulas
expressing the $N$ zeros $x_{n}$ in terms of the $N$ coefficients 
$y_n$ using radicals are
generally only available for $N\leq 4$.

\subsection{Definitions}

\label{subsec:def2}

The monic polynomial (\ref{pN}) is hereafter called \textit{peculiar} if it
has the \textit{peculiar} property that its $N$ zeros $x_{n}$ can be ordered
so as to coincide one-by-one with its $N$ coefficients $y_{m}$. We define $%
\mathcal{P}^{(N)}$ as the \textit{set} of all such peculiar polynomials. In
other words, $p_{N}(z)\in \mathcal{P}^{(N)}$ if and only if there exists a $%
\vec{y}$ such that 
\begin{equation}
p_{N}(z)=p_{N}(z;\vec{y},\vec{y})~.
\end{equation}%
This set, as we shall see, can be divided into \textit{non-empty\/}
subsets: one is ${\mathcal{P}_{0}}^{(N)}$, defined as the set of peculiar
polynomials having at least one coefficient equal to $0$. Another is ${%
\mathcal{P}_{1}}^{(N)}$, defined as the set of all peculiar polynomials
having at least one coefficient equal to $1$. Finally, we define the set of
the \textit{truly peculiar} polynomials $\mathcal{P}_{t}^{(N)}$ as 
\begin{equation}
{\mathcal{P}_{t}}^{(N)}=\mathcal{P}^{(N)}\backslash \left( \mathcal{P}%
_{0}^{(N)}\cap \mathcal{P}_{1}^{(N)}\right) ,
\end{equation}%
that is, the \textit{truly peculiar} polynomials are those \textit{peculiar}
polynomials having no coefficient (nor, of course, zero) which vanishes or
equals unity. The reason for this nomenclature is that, as we shall see
below, whenever one of the coefficients of a peculiar polynomial is either $0
$ or $1$, the problem of identifying it can be reduced to one involving a
lesser number of variables---i. e., polynomials of smaller degrees---whereas
the elements of ${\mathcal{P}_{t}^{(N)}}$ are the truly \textit{new\/}
polynomials of degree $N$ having the property of being \textit{peculiar}. As usual, for
any set $\mathcal{S}$, we denote by $|\mathcal{S}|$ the number of elements
of $\mathcal{S}$.

\section{Results}
\label{sec:results} In this section we state two results, a conjecture and a
third result which holds conditionally on the conjecture. We provide the
proofs in Section \ref{sec:proofs}. 

We begin by describing the structure of the set $\mathcal{P}^{(N)}$. First
note the following elementary fact: any $p_{N}(z)\in \mathcal{P}_{0}^{(N)}$
can be written as 
\begin{equation}
p_{N}(z)=zp_{N-1}(z)
\label{eq:corr}
\end{equation}%
where $p_{N-1}(z)\in \mathcal{P}^{(N-1)}$. There is therefore an elementary
one-to-one correspondence between $\mathcal{P}_{0}^{(N)}$ and $\mathcal{P}%
^{(N-1)}$. There moreover holds

\begin{proposition}
\label{prop1} For all $N\geq 3$, the set $\mathcal{P}^{(N)}$ can be divided
in the following way
\begin{equation}
\mathcal{P}^{(N)}=
\mathcal{P}_{t}^{(N)}\cup
\left(
\mathcal{P}_{1}^{(N)}\backslash \mathcal{P}_{0}^{(N)}
\right)
\cup
\mathcal{P}_{0}^{(N)},
\label{eq:disjoint}
\end{equation}
where the three sets $\mathcal{P}_{t}^{(N)}$, $\mathcal{P}_{1}^{(N)}\backslash \mathcal{P}_{0}^{(N)}$ and 
$\mathcal{P}_{0}^{(N)}$ are disjoint and non-empty
$\blacksquare $
\end{proposition}

\begin{remark}
Because of the elementary correspondence noted above, see (\ref{eq:corr}), 
between $\mathcal{P}_{0}^{(N)}$ and $\mathcal{P}^{(N-1)}$, we can proceed recursively to
divide $\mathcal{P}_{0}^{(N)}$ further by using \textbf{Proposition \ref{prop1}}
for $\mathcal{P}^{(N-1)}$. $\blacksquare $
\end{remark}

\begin{remark}
Note that \textbf{Proposition \ref{prop1}} fails when $N=2$, because, as we
shall see below, $\mathcal{P}_{t}^{(2)}$ is the empty set,
$\mathcal{P}_{t}^{(2)}=\emptyset $. $\blacksquare $
\end{remark}

Our second and main result concerns the number of peculiar polynomials of different
types.

\begin{proposition}
\label{prop2} For all $N\geq 3$:

\begin{subequations}
\begin{eqnarray}
\left|{\mathcal{P}^{(N)}} \right|&\leq& N!,
\label{ineq1}\\
\left|{\mathcal{P}_0^{(N)}} \right|&\leq&(N-1)!,
\label{ineq2}\\
\left|{\mathcal{P}_1^{(N)}} \right|&\leq&(N-1)!,
\label{ineq3}\\
\left|{\mathcal{P}^{(N)}}\backslash{\mathcal{P}_0^{(N)}} \right|&\leq&
(N-1)(N-1)!,
\label{ineq4}\\
\left|{\mathcal{P}_1^{(N)}}\backslash{\mathcal{P}_0^{(N)}}
\right|&\leq&(N-2)(N-2)!,
\label{ineq5}\\
\left|{\mathcal{P}_t^{(N)}} \right|&\leq& (N^2-3N+3)(N-2)!.
\label{ineq6}
\end{eqnarray}
\label{ineq}
\end{subequations}
In the following instances the above inequalities are strict: 
\begin{subequations}
\begin{eqnarray}
\left\vert {\mathcal{P}_{1}^{(4)}}\backslash {\mathcal{P}_{0}^{(4)}}%
\right\vert=3&<&4,
\label{strict1}\\
\left\vert \mathcal{P}_{0}^{(N+1)}\right\vert =\left\vert \mathcal{P}%
^{(N)}\right\vert&<&N!\qquad(N\geq 4),
\label{strict2}\\
\left\vert \mathcal{P}_{1}^{(N)}\right\vert&<&(N-1)!\qquad(N\geq 4)
.
\label{strict3}
\end{eqnarray}
\label{strict}
\end{subequations}
$ \blacksquare$
\end{proposition}

\begin{remark}
Note that the upper bounds (\ref{ineq2}, \ref{ineq5}, \ref{ineq6}) on the number of elements of the $3$ sets 
$\mathcal{P}_{0}^{(N)}$, ${\mathcal{P}_{1}^{(N)}}\backslash 
{\mathcal{P}_{0}^{(N)}}$ and ${\mathcal{P}_{t}^{(N)}}$ which build up $\mathcal{P}^{(N)}$
according to \textbf{Proposition \ref{prop1}} add up to $N!$:
\begin{equation}
\left\vert {\mathcal{P}_{0}^{(N)}}\right\vert +\left\vert {\mathcal{P}%
_{1}^{(N)}}\backslash {\mathcal{P}_{0}^{(N)}}\right\vert +\left\vert {%
\mathcal{P}_{t}^{(N)}}\right\vert \leq N!
\end{equation}
consistently with (\ref{eq:corr}) and (\ref{eq:disjoint}).
This implies that, if only one of these $3$ sets has strictly fewer elements
than what is given by the relevant upper bound given above, then the
inequality (\ref{ineq1}) for $\left\vert {\mathcal{P}^{(N)}}\right\vert $
will be strict, see (\ref{strict2}). Indeed, as follows
from the special case (\ref{strict1}), this always happens whenever $N\geq
4$, implying (\ref{strict2}). The validity of the inequality (\ref{strict3})
is implied by an analogous argument. It is thus seen that the second
and third strict inequalities (\ref{strict2}, \ref{strict3}) both follow from the first
strict inequality (\ref{strict1}). $\blacksquare $
\end{remark}

We now give the expressions for low-degree peculiar polynomials, which can
be obtained by solving the equations presented in \textbf{Section \ref%
{sec:proofs}} via such a program as Mathematica \cite{math}. Since the
polynomials of ${\mathcal{P}_{0}^{(N)}}$ can easily be obtained from the
polynomials belonging to the sets ${\mathcal{P}_{1}^{(M)}}\backslash 
\mathcal{P}_{0}^{(M)}$ and ${\mathcal{P}_{t}^{(M)}}$ with $M<N$, we only
give the polynomials belonging to the sets ${\mathcal{P}_{1}^{(N)}}%
\backslash {\mathcal{P}_{0}^{(N)}}$ and ${\mathcal{P}_{t}^{(N)}}$. For $%
N\leq 4$ these peculiar polynomials are as follows:

For $N=2$: ${\mathcal{P}_{t}^{(2)}}=\emptyset $ and there is 1
polynomial $z^{2}+z-2=\left( z-1\right) \left( z+2\right)$ belonging to
${\mathcal{P}%
_{1}^{(2)}}\backslash {\mathcal{P}_{0}^{(2)}}$.

For $N=3$: there is $1$ polynomial $z^{3}+z^{2}-z-1=\left( z-1\right) \left(
z+1\right) ^{2}\in {\mathcal{P}_{1}^{(3)}}\backslash {\mathcal{P}_{0}^{(3)}}$
and $3$ polynomials $z^{3}+y_{1}z^{2}+y_{2}z+y_{3}\in {\mathcal{P}_{t}^{(3)}}
$ where 
\begin{subequations}
\begin{eqnarray}
y_{1} &=&w, \\
y_{2} &=&-1-2w+2w^{3}, \\
y_{3} &=&-1-2w^{3},
\end{eqnarray}%
the number $w$ being one of the $3$ roots of the cubic equation 
$2w^{3}+2w^{2}-1=0$ (only one of which is \textit{real}, see above).
Note that this equation is {\em irreducible\/} over the rationals, a fact we shall 
make use of later. In particular, note that here and throughout in the following, 
whenever we speak of an  {\em irreducible\/} polynomial, we shall always implicitly 
mean that this should be understood over the rationals. 

For $N=4$: there are altogether $17$ polynomials in 
$ {\mathcal{P}^{(4)}}\backslash {\mathcal{P}_{0}^{(4)}}$, of which $3$
read as follows: $z^{4}+z^{3}+y_{2}z^{2}+y_{3}z+y_{4}\in {\mathcal{P}%
_{1}^{(4)}}\backslash {\mathcal{P}_{0}^{(4)}}$, where 
\end{subequations}
\begin{subequations}
\begin{eqnarray}
y_{2} &=&w, \\
y_{3} &=&-w^{3}-3w^{2}-3w-2, \\
y_{4} &=&w(w^{2}+3w+2),
\end{eqnarray}%
and $w$ is one of the $3$ roots of the {\em irreducible\/} cubic equation 
\begin{equation}
w^{3}+2w^{2}+w+1=0.
\end{equation}%
\label{eq:41minus0}
\end{subequations}
Again, note that only $1$ root of this cubic is \textit{real }(see above).

The remaining $14$ polynomials, 
none of which is \textit{real}, read 
$z^{4}+y_{1}z^{3}+y_{2}z^{2}+y_{3}z+y_{4}\in {\mathcal{P}_{t}^{(4)}}$ with 
\begin{subequations}
\begin{eqnarray}
y_{1} &=&w~, \\
y_{2} &=&\frac{1}{3301}\big(-5780-9301~w-15701~w^{2}-19444~w^{3}  \notag \\
&&\quad +15074~w^{4}+62196~w^{5}+79384~w^{6}+62708~w^{7}  \notag \\
&&\quad +7240~w^{8}-87856~w^{9}-157888~w^{10}-149344~w^{11}  \notag \\
&&\quad -79664~w^{12}-17888~w^{13}\big)~, \\
y_{3} &=&\frac{1}{3301}\big(2950-7255~w-24398~w^{2}-24629~w^{3}  \notag \\
&&\quad +30824~w^{4}+51850~w^{5}+60560~w^{6}+34348~w^{7}  \notag \\
&&\quad -35540~w^{8}-129696~w^{9}-116400~w^{10}-27936~w^{11}  \notag \\
&&\quad +78384~w^{12}+69808~w^{13}+30528~w^{14}-9903~y_{2}  \notag \\
&&\quad -9903~y_{2}^{2}-3301~y_{2}^{3}\big)~, \\
y_{4} &=&\frac{1}{3301}\big(-2950+653~w+24398~w^{2}+24629~w^{3}  \notag \\
&&\quad -30824~w^{4}-51850~w^{5}-60560~w^{6}-34348~w^{7}  \notag \\
&&\quad +35540~w^{8}+129696~w^{9}+116400~w^{10}+27936~w^{11}  \notag \\
&&\quad -78384~w^{12}-69808~w^{13}-30528~w^{14}+6602~y_{2}  \notag \\
&&\quad +9903~y_{2}^{2}+3301~y_{2}^{3}\big)~,
\end{eqnarray}%
where $w$ is one of the $14$ roots---\textit{all different} among themselves
and \textit{complex}, constituted of course by $7$ \textit{complex conjugate}
pairs---of the following {\em irreducible\/} polynomial equation of degree $14$:%
\begin{eqnarray}
&&1+3w+6w^{2}+6w^{3}+3w^{4}-12w^{5}-34w^{6}-44w^{7}  \notag \\
&&\quad -28w^{8}+4w^{9}+48w^{10}+80w^{11}+80w^{12}+48w^{13}+16w^{14}=0~.
\end{eqnarray}
\end{subequations}
This ends our treatment of the $N=4$ case.

Note the remarkable way these findings valid in the \textit{complex}
context extend those found by Stein in the \textit{real} case (as reported
in Section \ref{sec:def} above).
For $N>4$ the extension to the \textit{complex} case of Stein's findings
for the \textit{real} case is even more significant, see below, but the results
become too unwieldy to permit their \textit{explicit} display.

We have seen so far that $\mathcal{P}^{(N)}$ can be divided into simpler
subsets, each of which can be brought into one-to-one correspondence,
using (\ref{eq:corr}), with sets of the type $\mathcal{P}_{t}^{(M)}$ and $\mathcal{P}%
_{1}^{(M)}\backslash \mathcal{P}_{0}^{(M)}$ for $M\leq N$. We may ask
whether the subdivision can go further. That this will not happen for all $N$
is immediately clear from the results presented above for $2\leq N\leq 4$.

Indeed, for all these values of $N$, we find that the sets 
$\mathcal{P}_{t}^{(N)}$ and $\mathcal{P}_{1}^{(N)}\backslash \mathcal{P}_{0}^{(N)}$ 
are \textit{irreducible} over $\mathbb{Q}$ in the following sense: $y_{1}$ can, in all cases, be
expressed as the root of an appropriate polynomial $\Pi_1(y)$ with rational coefficients,
{\em irreducible\/} over $\mathbb{Q}$.   
The $y_{k}$ for $2\leq k\leq N$ are then given as
polynomials of $y_{1}$. Clearly, the peculiar role of $y_1$ is only apparent: 
for any $k$, with $1\leq k\leq N$, the $y_{l}$ with $l\neq k$ can then also
be expressed in terms of $y_{k}$, which in turn is then the root of an
appropriate \textit{irreducible} polynomial $\Pi_k(y)$.
Clearly, if the polynomial $\Pi_1(y)$ which defines a given set of polynomials, 
were reducible, we could factorize it into polynomials with integer coefficients. 
This would then allow to divide this subset further according to whether
$y_1$ is a zero of one or the other of the factors. Since the $\Pi_1(y)$ in all the
cases described above, are in fact irreducible, such an additional reduction of
the subsets described above is not possible.

In the following, we shall show that all elements of 
$\mathcal{P}_{1}^{(N)}\backslash \mathcal{P}_{0}^{(N)}$ satisfy the system of equations 
\begin{subequations}
\begin{eqnarray}
y_{2}+\ldots +y_{N} &=&-2, \\
\sigma _{m}(1,y_{2},\ldots ,y_{N}) &=&(-1)^{m}y_{m},\qquad (2\leq m\leq N-2),
\\
y_{2}\cdot \ldots \cdot y_{N-1} &=&(-1)^{N}.
\end{eqnarray}
\label{eq:1minus0}
\end{subequations}
Similarly, we shall see that the elements of $\mathcal{P}_{t}^{(N)}$ all
satisfy the system of equations: 
\begin{subequations}
\begin{eqnarray}
&&\sigma _{m}(\vec{y})=(-1)^{m}y_{m}\qquad (1\leq m\leq N-2), \\
&&y_{1}\cdot \ldots \cdot y_{N-1}=(-1)^{N}, \\
&&2y_{1}\left( \sum_{l=0}^{N-2}y_{1}^{l}\right) +\sum_{k=1}^{N-1}y_{k}\left(
\sum_{l=0}^{N-(k+1)}y_{1}^{l}\right) =0.
\end{eqnarray}
\label{eq:true}
\end{subequations}
From standard elimination theory, as described, for example, in \cite{Bezout}, one may show that 
the solutions of equations such as (\ref{eq:1minus0}) and (\ref{eq:true}), can be described by
saying that $y_1$ is the solution of a polynomial equation, whereas the $y_k$ for $2\leq k\leq N$
are expressed via a polynomial relation connecting $y_k$ and $y_1$ The differences with the characterizations 
above are twofold: the polynomial need not be irreducible and the connection between $y_k$ and $y_1$
is a general polynomial equation, whereas above we had noticed that  $y_k$ 
can be expressed as a polynomial expression of $y_1$. 

We now formulate the

\begin{conjecture}
For all $N\geq 5$, upon reducing  (via standard elimination theory, as above)
the systems of equations (\ref{eq:1minus0}) and (\ref{eq:true}),
$y_1$ can be expressed as the zero of a polynomial $\Pi_1(y)$,
irreducible, and the $y_k$'s
can be expressed as polynomials of $y_1$. 
$\blacksquare $
\end{conjecture}

\begin{remark}
Note that, for $N=4$, the total set of polynomials belonging to $\mathcal{P}%
_{1}^{(4)}\backslash \mathcal{P}_{0}^{(4)}$ can be characterized by an
irreducible polynomial as stated above, see (\ref{eq:1minus0}). On the other hand, the
{\bf Conjecture} does not hold in this case, because it is found that then
a solution of (\ref{eq:1minus0}) with $y_4=0$ exists. Indeed, this is the
exception which causes all the special cases identified above (in the last
part of \textbf{Proposition \ref{prop2}}). $\blacksquare $
\end{remark}

There then holds the following result:

\begin{proposition}
\label{prop3}
If the \textbf{Conjecture  } holds, then inequalities (\ref{ineq1}, \ref{ineq2}, \ref{ineq3}) 
of \textbf{Proposition \ref{prop2}} hold as equalities for 
$N\geq 5$, so that the cases of strict inequality (\ref{strict})---as
stated in the last part of \textbf{Proposition \ref{prop2}}---are the only
possible ones. $\blacksquare $
\end{proposition}

We consider the \textbf{Conjecture  } plausible but we recognize that the arguments
for its validity are so far limited to the cases we tested numerically via
Mathematica \cite{math}: $N\leq 7$ for (\ref{eq:1minus0}) and $N\leq 6$ for (\ref%
{eq:true}).

\section{Proofs}
\label{sec:proofs}

The coefficients $y_{n}$ of a peculiar polynomial $p_{N}(z)\in \mathcal{P}%
^{(N)}$ clearly satisfy the set of equations 
\begin{equation}
\sigma _{m}(y_{1},\ldots ,y_{N})=(-1)^{m}y_{m},\qquad (1\leq m\leq N).
\label{eq:1}
\end{equation}%
Here and below $\sigma _{m}(y_{1},\ldots ,y_{N})$ is of course defined by (%
\ref{sigmam}) with the $N$ zeros $x_{n}$ replaced by the $N$ coefficients $%
y_{m}$.These equations can be reformulated so as to be equations in
projective space, as follows 
\begin{equation}
\sigma _{m}(y_{1},\ldots ,y_{N})=(-1)^{m}y_{0}^{m-1}y_{m},\qquad (1\leq
m\leq N).  \label{eq:2}
\end{equation}%
To find the full number of solutions of (\ref{eq:2}) when counted with the
appropriate multiplicities, we apply B\'{e}zout's theorem
(see, for instance, \cite{Bezout} for a statement and an elementary proof).
Since, as it is readily verified, the only solution with $y_{0}=0$ is the
trivial solution where all $y_{n}$ vanish, the solutions of (\ref{eq:2}) 
are all solutions of (\ref{eq:1}). The number of solutions of 
(\ref{eq:1}) counted in the same manner is therefore the
product of the degrees of all these equations (\ref{eq:1}), that is, $N!$.
Since all elements of $\mathcal{P}^{(N)}$ satisfy (\ref{eq:1}) and since multiplicities
are always larger or equal to one, the validity
of inequality (\ref{ineq1}) of \textbf{Proposition \ref{prop2}} is
thereby demonstrated. 

Note in passing that the coefficients $y_{n}$ of a peculiar polynomial $%
p_{N}(z)$ must, of course, satisfy the $N$ algebraic equations 
\begin{equation}
p_{N}(y_{m})=y_{m}^{N}+\sum_{r=1}^{N}y_{r}y_{m}^{N-r}=0,\qquad (1\leq m\leq
N).  \label{eq:3}
\end{equation}%
However, contrary to (\ref{eq:2}), these equations do not guarantee that the
corresponding polynomial be \textit{peculiar}, because they do not imply
that \textit{all\/} zeros of the polynomial $p_{N}(z)$ are coefficients. A 
counterexample is thus given by the polynomial $p_{2}(z)=z^{2}-z/2-1/2$, 
the coefficients of which satisfy (\ref{eq:3}), but which is \textit{not
peculiar}, because $z=1$ is a zero of $p_{2}(z)$ without being one of its
coefficients. For future use, we point out the projective form of (\ref{eq:3}%
) 
\begin{equation}
y_{m}^{N}+\sum_{r=1}^{N}y_{0}^{r-1}y_{r}y_{m}^{N-r}=0,\qquad (1\leq m\leq N).
\label{eq:3p}
\end{equation}

Let us now consider the set $\mathcal{P}_{0}^{(N)},$ i. e. the case in which
one of the coefficients \textit{vanishes}, say $y_{k}=0$. It then follows
from (\ref{eq:2}) with $m=N$ that $y_{N}$ vanishes, $y_{N}=0$. From this
follows that (\ref{eq:2}) for $m=N$ is an identity, and can thus be
discarded. The system (\ref{eq:2}) then becomes a system of $N-1$ equations
in the $N-1$ unknowns, $y_{1},\ldots ,y_{N-1}$, to which we may again apply B%
\'{e}zout's theorem. It thus follows that, again counting multiplicities,
there are $(N-1)!$ solutions to this system of equations. Since each
peculiar polynomial $p_{N}(z)\in \mathcal{P}_{0}^{(N)}$ satisfies this
system of equations, and since multiplicities are always larger than or
equal to one, this proves inequality (\ref{ineq2}) of \textbf{%
Proposition \ref{prop2}}. Note that this result could also have been
obtained from the simple remark 
\begin{equation}
\left\vert \mathcal{P}_{0}^{(N)}\right\vert =\left\vert \mathcal{P}%
^{(N-1)}\right\vert 
\end{equation}%
and inequality (\ref{ineq1}) applied with $N$ replaced by $N-1$.

Let us next consider the set $\mathcal{P}_{1}^{(N)},$ i. e. the case in
which one of the coefficients equals \textit{unity}, say $y_{k}=1$, which in the projective 
formulation amounts to saying $y_k=y_0$. 
We first show that then $y_{1}=y_0$. To this end, we consider (\ref{eq:3p}) for the case
in which $m=k$. It then follows from $y_{k}=y_0$ that 
\begin{equation}
y_0+\sum_{r=1}^{N}y_{r}=0.  \label{eq:5}
\end{equation}%
Comparing this with (\ref{eq:2}) for $m=1$ we immediately obtain that $%
y_{1}=y_0$. 

We may now replace $y_{1}$ by $y_0$ in all the equations (\ref{eq:2}). This
yields 
\begin{subequations}
\begin{eqnarray}
\sigma _1(y_0,y_{2},\ldots ,y_{N})&=&-y_{0}
\label{eq:6a}\\
\sigma _{m}(y_0,y_{2},\ldots ,y_{N})&=&(-1)^{m}y_0^{m-1}y_{m}\qquad (2\leq m\leq N).
\label{eq:6b}
\end{eqnarray}%
\label{eq:6}
\end{subequations}
This system of $N$ algebraic equations can clearly be rewritten as follows: 
\begin{subequations}
\label{eq:7}
\begin{eqnarray}
&&\sigma _{1}(0,y_{2},\ldots ,y_{N})=-2y_0,  \label{eq:7a} \\
&&(-1)^{m}\left[ 
y_0
\sigma _{m-1}(0,y_{2},\ldots ,y_{N})+
\sigma_{m}(0,y_{2},\ldots ,y_{N})
\right] =y_{0}^{m-1}y_{m},  \label{eq:7b} \\
&&(-1)^{N}\sigma _{N-1}(0,y_{2},\ldots ,y_{N})=y_{0}^{N-2}y_{N},
\label{eq:7c}
\end{eqnarray}
\end{subequations}
where, in (\ref{eq:7b}), $2\leq m\leq N-2$. We now combine equations 
(\ref{eq:7b}) with each other for all values of $m$ and, via (\ref{eq:7a}), we
obtain the last, (\ref{eq:7c}); which is thus seen to be \textit{superfluous}. 
The equations satisfied by the coefficients of a polynomial belonging to $%
\mathcal{P}_{1}^{(N)}$ thus reduce to a set of $N-1$ equations for the $N-1$
unknowns $y_{2},\ldots ,y_{N}$. We therefore have described ${\mathcal{P}%
_{1}^{(N)}}$ by $N-1$ equations, the degrees of which have a product of $%
(N-1)!$. The coefficients of a polynomial $p_{N}(z)\in \mathcal{P}_{1}^{(N)}$
are thus characterized by the fact that they satisfy a system of equations
which has $(N-1)!$ solutions counted with multiplicity. It thus follows, as
stated by inequality (\ref{ineq3}) of \textbf{Proposition \ref{prop2}},
that the total number of such polynomials cannot exceed $(N-1)!$.

Note that we have shown that the sets $\mathcal{P}^{(N)}$, $\mathcal{P}%
_{0}^{(N)}$, and $\mathcal{P}_{1}^{(N)}$ correspond \textit{exactly}\/ to the
solutions of some appropriate system of algebraic equations. That is, a
polynomial belongs to one of these sets if and only if its coefficients
satisfy the corresponding set of equations. The difference between the
number of elements of these sets and their upper bounds reported in \textbf{%
Proposition \ref{prop2}} can thus \textit{only\/} arise from the existence of
multiple solutions in the equations derived above for each one of these $3$
sets. This will \textit{not hold any more }for the sets we consider in the
following.

Since we have, up to now, only obtained upper bounds for the cardinalities
of ${\mathcal{P}^{(N)}}$, ${\mathcal{P}_{0}^{(N)}}$ and ${\mathcal{P}^{(N)}}$, 
there does not follow any non-trivial estimate either from below or from
above, say, for 
$\left\vert {\mathcal{P}^{(N)}}\backslash {\mathcal{P}_{0}^{(N)}}\right\vert $. To 
obtain these, we proceed to determine equations
which the elements of such sets must satisfy. These will always provide an upper bound
on the number of elements in the set. However, we will not generally be able
to show that all the solutions of the equations belong to the set, but only
the converse: all elements of the set do satisfy the equations.

The simplest example is given by ${\mathcal{P}^{(N)}}\backslash {\mathcal{P}_{0}^{(N)}}$: 
in this case we may simplify the equation (\ref{eq:2}) to 
\begin{subequations}
\begin{eqnarray}
\sigma _{m}(y_{1},\ldots ,y_{N})&=&(-1)^{m}y_{0}^{m-1}y_{m},\qquad (1\leq
m\leq N-1)
 \label{eq:8a}\\
y_{1}\cdot \ldots \cdot y_{N-1}&=&(-1)^{N}y_0^{N-1}. \label{eq:8b}
\end{eqnarray}
\label{eq:8}
\end{subequations}
(\ref{eq:8b}) was obtained by dividing both sides of the equation (\ref{eq:2}) corresponding to
$m=N$ by $y_{N}$, which
requires, but does \textit{not imply}, $y_{N}\neq 0$. The degree of this
equation has thus decreased by one. Via B\'{e}zout's theorem we may then
deduce that the corresponding system of equations has exactly $(N-1)(N-1)!$
solutions counted with multiplicity, and hence, as above, that $\left\vert {%
\mathcal{P}^{(N)}}\backslash {\mathcal{P}_{0}^{(N)}}\right\vert \leq
(N-1)(N-1)!$, as stated by inequality (\ref{ineq4}) of \textbf{%
Proposition~\ref{prop2}}.

Note that one of the special cases in which the inequalities are strict, see (\ref{strict1}),
arises precisely from such an instance: when 
$N=4$, the set of equations (\ref{eq:8})
has among its solutions the following values of 
$y_{n}$: $y_{1}=1$, $y_{2}=-1$, $y_{3}=-1$ and $y_{4}=0$. There is thus an
element of ${\mathcal{P}_{0}^{(4)}}$ among the solutions of the above
equations. Note that this immediately implies that the corresponding zero is
a double zero of the set of equations (\ref{eq:2}), so that inequality
(\ref{ineq1}) of \textbf{Proposition \ref{prop2}} is then also strict.

Similarly, we may consider the case of ${\mathcal{P}_{1}^{(N)}}\backslash 
\mathcal{P}_{0}^{(N)}$. The set $\mathcal{P}_{1}^{(N)}$, as we have seen, is
described by (\ref{eq:7}) of which the
final one is superfluous. However, we can equally well discard the second
equation (\ref{eq:7b}) corresponding to $m=N-1$. If we then
use the assumption $y_{N}=0$ in order to divide by $y_{N}$ in the final
equation (\ref{eq:7c}), we are led to a system of $N-2$
equations, the orders of which have a product of $(N-2)!$, complemented by
one final equation of order $N-2$. These correspond exactly to (\ref{eq:1minus0}) . 
Since all elements of ${\mathcal{P}_{1}^{(N)}}\backslash 
\mathcal{P}_{0}^{(N)}$ satisfy these equations, the result claimed in
inequality (\ref{ineq5}) of \textbf{Proposition \ref{prop2}} is
shown.

To estimate $\left\vert \mathcal{P}_{t}^{(N)}\right\vert $, we proceed as
follows: first, we note that the elements of ${\mathcal{P}^{(N)}}\backslash {%
\mathcal{P}_{0}^{(N)}}$ satisfy the equations 
\begin{subequations}
\begin{eqnarray}
\sigma _{m}(y_{1},\ldots ,y_{N}) &=&(-1)^{m}y_{0}^{m-1}y_{m},\qquad (1\leq
m\leq N-1),
 \label{eq:13a}\\
y_{1}\cdot \ldots \cdot y_{N-1} &=&(-1)^{N}y_{0}^{N-1}.
\label{eq:13b}
\end{eqnarray}
\label{eq:13}
\end{subequations}
Let us now ask, within this set, how many elements satisfy $y_{1}=y_0$. As
above, we show that these solutions satisfy the equations: 

\begin{subequations}
\begin{eqnarray}
\sigma _1(y_{0},y_2,\ldots ,y_{N}) &=&-y_0,
\label{eq:14a}\\
\sigma _{m}(y_{0},y_2,\ldots ,y_{N}) &=&(-1)^{m}y_{0}^{m-1}y_{m},\qquad (2¼¼\leq
m\leq N-2), 
\label{eq:14b}\\
y_{2}\cdot \ldots \cdot y_{N-1} &=&(-1)^{N}y_{0}^{N-2}.
\label{eq:14c}
\end{eqnarray}
\label{eq:14}
\end{subequations}
Note that, here as in (\ref{eq:7}), we show that the
equation for $m=N-1$ follows from the others and can thus be discarded. From
B\'{e}zout's theorem follows that (\ref{eq:13}) have $%
(N-1)^{2}(N-2)!$ solutions, whereas (\ref{eq:14}) have $(N-2)(N-2)!$
solutions, both counted according to multiplicities. Now the solutions of (%
\ref{eq:14}) are a subset of those of (\ref{eq:13}). They thus have at most
the same multiplicities. It thus follows that $\left\vert {\mathcal{P}%
_{t}^{(N)}}\right\vert \leq \lbrack (N-1)^{2}-(N-2)](N-2)!$, or in other
words ${\mathcal{P}_{t}^{(N)}}\leq (N^{2}-3N+3)(N-2)!$ as stated in
inequality (\ref{ineq6}) of \textbf{Proposition \ref{prop2}}. 

We can also deduce that all elements of $\mathcal{P}_t^{(N)}$ satisfy 
(\ref{eq:true}): to this end we perform the following operations on (\ref{eq:1}):
divide the equation of (\ref{eq:1}) corresponding to $m=N$ by $y_N$, using
the fact that $y_N\neq0$. Then replace the equation corresponding to $m=N-1$
by the equation (\ref{eq:3p}) for $m=1$. Finally, subtract equation 
(\ref{eq:1}) for $m=1$ from that last equation, which may then be divided by $%
y_1-1$. The resulting equations are then exactly (\ref{eq:true}). Note that
the projective form of
these equations, that is, the equations (\ref{eq:true}) modified so as to be homogeneous
in the $y_m$'s via the introduction of the parameter $y_0$,
have a number of solutions ``at infinity'', that is,
corresponding to $y_0=0$. Since B\'ezout's theorem yields the number of solutions for the projective form 
of the equations, it follows that the number of solutions of (\ref{eq:true}) is strictly less than the
number obtained via B\'ezout's theorem, namely $(N-1)(N-1)!$, as indeed corresponds
to the statement of (\ref{ineq6}).

Finally, we need to prove \textbf{Proposition \ref{prop3}}. Assuming the
\textbf{Conjecture  }, we see that neither (\ref{eq:1minus0}) nor (\ref%
{eq:true}) can have multiple solutions, since an irreducible polynomial does
not have multiple zeros. It remains to see, then, that neither 
(\ref{eq:1minus0}) nor (\ref{eq:true}) can have solutions which do not correspond
to elements of $\mathcal{P}_{1}^{(N)}\backslash \mathcal{P}_{0}^{(N)}$, or
elements of $\mathcal{P}_{t}^{(N)}$ respectively. Let us begin with the
first case, the second being quite similar.

Let $(y_2,\ldots,y_N)$ be a solution of (\ref{eq:1minus0}). Then it is clear
that the polynomial $z^N+z^{N-1}+\sum_{k=2}^Ny_kz^{N-k}$ is peculiar, and
has a coefficient $y_1=1$. It thus merely remains to show that $y_N\neq0$.
From (\ref{eq:1minus0}) and $y_N=0$ follows, combining the equation for $%
m=N-2$ and the last equation, that $y_{N-1}=-1$. According to the
conjecture, however, the solution set of (\ref{eq:1minus0}) can be described
by stating that $y_{N-1}$ is the root of an appropriate polynomial,
irreducible over $\mathbb{Q}$, and that the $y_k$ with $k\neq N-1$ are
expressed by polynomials in $y_{N-1}$. Now the very fact that $y_{N-1}=-1$
contradicts irreducibility. One shows quite similarly that (\ref{eq:true})
can have no solutions with $y_N=0$, again because in that case $y_{N-1}=-1$.
It therefore follows that $\left|\mathcal{P}_1^{(N)}\backslash\mathcal{P%
}_0^{(N)}\right|$ and $\left|\mathcal{P}_t^{(N)}\right|$ are equal to $%
(N-2)(N-2)!$ and $(N^2-3N+3)(N-2)!$, respectively.

\section{Conclusions}
\label{sec:conclusions}

Summarizing, we have shown that peculiar polynomials, 
defined as those polynomials which
remain identical when we replace their coefficients by (an appropriate permutation of) 
their zeros, exist for all degrees $N$. The set of such polynomials can be divided
into three disjoint sets, namely those which have one coefficient equal to 0, those which have no 
coefficient equal to 0, but at least one equal to 1, and finally those which have no coefficients
equal to either 1 or 0. The first of these three sets can further be divided, since each of its 
elements corresponds to a peculiar polynomial of degree $N-1$.
We further show that, if we consider complex solutions, the above sets are all non-empty for all $N\geq3$.
This is in striking contrast to the 
corresponding result of Stein \cite{Stein1966} stating that no peculiar polynomials
with real coefficients all different from 0,
can exist for $N>4$. We have finally given upper bounds for
the number of elements of these various constituent sets, which are
presumably quite close to the actual values. Under an additional conjecture,
we show that the cases of strict inequality numbered in (\ref{strict})
are actually the only ones, so that for all $N\geq5$ the inequalities 
(\ref{ineq4}, \ref{ineq5}, \ref{ineq6}) hold as equalities. 

\section*{Acknowledgment.}
These findings were obtaining during the Gathering of Scientists held in
November--December 2016 at the Centro Internacional de Ciencias (CIC) in
Cuernavaca, Mexico. It is a pleasure to thank the participants for
interesting discussions, and in particular Decio Levi for making available
to us a copy of \cite{Ulam1960}. FL is further indebted to CONACyT grant
Ciencias B\'asicas 254515 and UNAM---PAPIIT---DGAPA grant IN103017
for financial support.

\end{document}